\documentclass[sigconf,screen]{acmart}

\AtBeginDocument{%
  }

\setcopyright{rightsretained}
\acmDOI{10.1145/3650212.3652143}
\acmYear{2024}
\copyrightyear{2024}
\acmSubmissionID{issta24main-p1189-p}
\acmISBN{979-8-4007-0612-7/24/09}
\acmConference[ISSTA '24]{Proceedings of the 33rd ACM SIGSOFT International Symposium on Software Testing and Analysis}{September 16--20, 2024}{Vienna, Austria}
\acmBooktitle{Proceedings of the 33rd ACM SIGSOFT International Symposium on Software Testing and Analysis (ISSTA '24), September 16--20, 2024, Vienna, Austria}
\received{16-DEC-2023}
\received[accepted]{2024-03-02}

\usepackage{listings}
\usepackage{xcolor}
\usepackage{multirow}
\usepackage{graphics}
\usepackage{tabularray}
\usepackage{tabularx}
\usepackage{pifont}
\usepackage{amsmath,bm}
\usepackage{tcolorbox}
\usepackage{mdframed,lipsum}
\usepackage{subcaption}
\usepackage{microtype}

\usepackage{xparse}
\usepackage{xspace}
\usepackage{comment}
\usepackage{soul}
\usepackage{colortbl}
\usepackage{pdfpages}
\usepackage[T1]{fontenc}

\DeclareTextFontCommand{\mytexttt}{\ttfamily\hyphenchar\font=45\relax}

\newcolumntype{P}[1]{>{\centering\arraybackslash}p{#1}}

\definecolor{candyapplered}{rgb}{1.0, 0.03, 0.0}
\definecolor{green_a}{HTML}{1b7837}
\definecolor{green_b}{HTML}{7fbf7b}
\definecolor{green_c}{HTML}{d9f0d3}
\definecolor{green_d}{HTML}{f6e8c3}
\definecolor{lightmagenta}{HTML}{E1BEE7}
\definecolor{lightpurple}{HTML}{C5CAE9}
\definecolor{lightblue}{HTML}{B3E5FC}
\definecolor{lightcyan}{HTML}{B2DFDB}

\usepackage{enumitem, kantlipsum}
\usepackage{filecontents,lipsum}
\usepackage{graphicx}
\usepackage{multirow}
\usepackage[linesnumbered,ruled,vlined]{algorithm2e}

\SetCommentSty{mycommfont}

\SetKwInput{KwInput}{Input}                
\SetKwInput{KwOutput}{Output}              

\newcommand{\Tool}{\textsc{Adopter}\xspace}

\newcommand{\CodeIn}[1]{{\mytexttt{\small #1}}}

\newcommand{\Comment}[1]{}

\newcommand{\distance}{8pt}
\newcounter{definition}

\newcounter{finding}
\newenvironment{finding}[1]
{
    \refstepcounter{finding}
	\begin{mdframed}[
    	nobreak=true,
    	linecolor=black,
    	roundcorner=12pt,
    	backgroundcolor=gray!01,
    	linewidth=0.5pt,
    	leftmargin=0.5em,
    	rightmargin=0.5em,
    	topline=true,
    	bottomline=true,
    	frametitlerule=true,
    	frametitlebackgroundcolor=gray!20,
    	frametitlerulecolor=gray,
    	frametitle=Answer to RQ\arabic{finding},
    	frametitleaboveskip=0.3em,
    	frametitlebelowskip=0.2em,
    	skipabove=12pt
	]
}
{
    \end{mdframed}
    \vspace{1em}
}

\begin{document}

\title[Automated Deep Learning Optimization via DSL-Based Source Code Transformation]{Automated Deep Learning Optimization via \\ DSL-Based Source Code Transformation}

\author{Ruixin Wang}
\email{ruixinw@purdue.edu}
\orcid{0009-0002-9298-0925}
\affiliation{%
  \institution{Purdue University}
  \city{West Lafayette}
  \state{IN}
  \country{USA}
}

\author{Minghai Lu}
\email{lu1074@purdue.edu}
\orcid{0009-0001-0136-3204}
\affiliation{%
  \institution{Purdue University}
  \city{West Lafayette}
  \state{IN}
  \country{USA}
}

\author{Cody Hao Yu}
\email{cody@boson.ai}
\orcid{0000-0002-9298-6254}
\affiliation{%
  \institution{BosonAI}
  \city{Santa Clara}
  \state{CA}
  \country{USA}
}

\author{Yi-Hsiang Lai}
\email{yihsian@amazon.com}
\orcid{0000-0002-2358-805X}
\affiliation{%
  \institution{Amazon Web Services}
  \city{Seattle}
  \state{WA}
  \country{USA}
}

\author{Tianyi Zhang}
\email{tianyi@purdue.edu}
\orcid{0000-0002-5468-9347}
\affiliation{%
  \institution{Purdue University}
  \city{West Lafayette}
  \state{IN}
  \country{USA}
}



\begin{abstract}
As deep learning models become increasingly bigger and more complex, it is critical to improve model training and inference efficiency. 
Though a variety of highly optimized libraries and packages (known as {\em DL kernels}) have been developed, it is tedious and time-consuming to figure out which kernel to use, where to use, and how to use them correctly.  
To address this challenge, we propose an {A}utomated {D}eep learning {OPT}imization approach called \Tool. We design a Domain-Specific Language (DSL) to represent DL model architectures and leverage this DSL to specify model transformation rules required to integrate a DL kernel into a model. Given the source code of a DL model and the transformation rules for a set of kernels, \Tool first performs inter-procedural analysis to identify and express the model architecture in our DSL. Then, \Tool performs scope analysis and sub-sequence matching to identify locations in the model architecture where the transformation rules can be applied. Finally, \Tool proposes a synthesis-based code transformation method to apply the transformation rule. We curated a benchmark with {199} models from Hugging Face and a diverse set of DL kernels. We found that, compared to a state-of-the-art automated code transformation technique, \Tool helps improve the precision and recall by 3\% and 56\%, respectively. An in-depth analysis of 9 models revealed that on average, \Tool improved the training speed by 22.7\% while decreasing the GPU memory usage by 10.5\%.  
\end{abstract}



\maketitle


\section{Introduction}
Deep learning (DL) models have experienced rapid growth in many domains, e.g., computer vision~\cite{dosovitskiy_image_2021, he_deep_2015, zhang_xformer_2023}, natural language processing~\cite{raffel_exploring_2020, vaswani_attention_2023}, robotics ~\cite{kumra_robotic_2017, mahler_dex-net_2017}, bioinformatics~\cite{min_deep_2016, bacciu_bioinformatics_2018, karim_deep_2021, lan_survey_2018}, etc. As these models increase in size and complexity, improving their computational efficiency has become an emerging challenge.
To address this challenge, a variety of highly optimized libraries and packages, which are known as {\em DL kernels}, have been developed to accelerate model training and inference, such as Nvidia's Apex~\cite{apex}, Microsoft's Deepspeed~\cite{deepspeed}, and Meta's xFormers~\cite{xsformer}.  

While optimized DL kernels help address computational demands, effectively leveraging them poses challenges for developers.
First, since optimized kernels are dispersed across different platforms and are constantly evolving, developers face challenges in searching through forums, blogs, and code repositories to find appropriate kernels optimized for their own models. 
Even after finding a suitable kernel, developers still need to delve into documentations and tutorials to determine which parts of their models can be replaced with the kernel and how to change their code to use the kernel. 
This manual integration process requires extensive manual effort and is error-prone. 

To address this challenge, we propose \Tool, an automated approach that automatically integrates DL kernels into developers' models. Since model implementations are full of intricate syntax details and extraneous statements, it is challenging to directly match potential optimization locations to apply a kernel. Thus, we design a Domain-Specific Language (DSL) that abstracts program source code to represent model architectures as sequences of DL layers and tensor operations. We further leverage this DSL to specify model transformation rules required to integrate a DL kernel into a model. Specifically, we defined nine optimization rules for a diverse set of DL kernels collected from GitHub~\cite{epoi, noauthor_triton_2023, noauthor_slapo_2023}, HuggingFace~\cite{huggingface}, and research artifacts~\cite{noauthor_slapo_2023}.

Given the source code of a DL model and the pre-defined transformation rules, \Tool first performs an inter-procedural control flow analysis to extract the model architecture and represent it in our DSL. Then, \Tool performs scope analysis and sub-sequence matching to identify locations in the model architecture where the transformation rules can be applied. Finally, for each matched location where the transformation rules can be applied, \Tool employs a synthesis-based code transformation method to integrate the optimized kernel. 

Compared with existing automated program transformation  techniques~\cite{meng_systematic_2011, meng_lase_2013, dilhara_pyevolve_2023},  
\Tool can more accurately identify optimization opportunities while accounting for the syntactic variations and extraneous code in model implementations by abstracting a model implementation into a DSL and performing pattern matching on the DSL representation.  
Furthermore, the synthesis-based transformation approach in \Tool supports more complex transformations, e.g., merging a statement before an if statement with a statement within a statement.  


We evaluate the effectiveness of \Tool on a benchmark with 199 DL models from Hugging Face. Our results show that compared to PyEvolve~\cite{dilhara_pyevolve_2023}, a state-of-the-art code transformation approach, \Tool improves the precision and recall by 4.4\% and 44.9\%, respectively. Furthermore, we performed an in-depth analysis of 9 DL models and measured whether the optimizations applied by \Tool indeed improved computation efficiency. Our results show that the training speed of these models was increased by 15.3\% on average, while the GPU memory usage was decreased by 6.8\% on average. 
Finally, an ablation study shows that the inter-procedural control-flow analysis and the scope analysis methods leveraged by our approach can significantly improve the effectiveness of \Tool.

Overall, our work makes the following contributions:

\begin{itemize}[leftmargin=0.6cm]
 \item We introduce an expressive domain-specific language to describe model architectures and specify optimization rules. 
 \item We propose \Tool, a novel approach that automatically integrates DL kernels into model implementations based on the DSL. Our source code has been made publicly available on Zenodo~\cite{zenodo_artifact} and GitHub\footnote{\label{note1}\href{https://github.com/ailen-wrx/Adopter}{https://github.com/ailen-wrx/Adopter}} to foster future research. 
 \item We create and open-source a benchmark with a diverse set of DL optimization rules and 199 Hugging Face models. Our benchmark is useful to evaluate the effectiveness of automatically optimizing DL models on source code level.
 \item Our experiments on the benchmark show that compared to a state-of-the-art automatic code transformation technique, \Tool helps improve the precision and recall by 3\% and 56\%, respectively; our optimization rules help improve the model performance by increasing the {\em Samples pre Second} metric by 22.7\% and decreasing the {\em per GPU memory} metric by 10.5\%. 

\end{itemize}

The rest of the paper is organized as follows. Section \ref{sec:MotivatingExample} motivates the problem with a real-world example. Section \ref{sec:Methodology} describes the automated approach. Section \ref{sec:Evaluation} describes the experiments conducted to measure the effectiveness of our approach. Section \ref{sec:Discussion} discusses the potential of future research and the threats to the validity. Section \ref{sec:RelatedWork} discusses related work, and Section \ref{sec:Conclusion} concludes the paper.

\section{Motivating Example}
\label{sec:MotivatingExample}
\definecolor{codegreen}{rgb}{0.13,0.41,0.24}
\definecolor{codegray}{rgb}{0.5,0.5,0.5}
\definecolor{codepurple}{rgb}{0.52,0.03,0.40}
\definecolor{backcolour}{rgb}{1.0,1.0,1.0}

\definecolor{patchgreen}{HTML}{ccffcc}
\definecolor{patchred}{HTML}{ffcccc}
\definecolor{matchyellow}{HTML}{FCF3CF}

\newcommand{\addgreen}{\makebox[0pt][l]{\color{patchgreen}\rule[-1ex]{\linewidth}{3ex}}}
\newcommand{\addred}{\makebox[0pt][l]{\color{patchred}\rule[-1ex]{\linewidth}{3ex}}}
\newcommand{\addyellow}{\makebox[0pt][l]{\color{matchyellow}\rule[-1ex]{\linewidth}{3ex}}}

\lstdefinestyle{code}{ %
  columns=fixed,
  basewidth=0.53em,
  basicstyle=\ttfamily\footnotesize,
  backgroundcolor=\color{backcolour},
  commentstyle=\color{codegreen},
  keywordstyle=\color{codepurple},
  numberstyle=\tiny\color{codegray},
  stringstyle=\color{magenta},
  numbers=left,
  numbersep=5pt,
  xleftmargin=0em,
  xrightmargin=0em,
  breaklines=true,
  frame=none,
  framexleftmargin=0em,
  framexrightmargin=0em,
  rulesepcolor=\color{gray},
  literate=
}

\lstset{style=code}

\let\origthelstnumber\thelstnumber
\makeatletter
\newcommand*\Suppressnumber{%
  \lst@AddToHook{OnNewLine}{%
    \let\thelstnumber\relax%
     \advance\c@lstnumber-\@ne\relax%
    }%
}

\newcommand*\Reactivatenumber{%
  \lst@AddToHook{OnNewLine}{%
   \let\thelstnumber\origthelstnumber%
   \advance\c@lstnumber\@ne\relax}%
}

\makeatother
{\begin{figure}[hbtp]
\begin{lstlisting}[language=Python, firstnumber=1, escapechar=|]
 class BertSelfOutput(nn.Module):
   def __init__(self, config):
     super().__init__()
     self.dense = nn.Linear(config.hidden_size,config.hidden_size)
|\addred| -   self.LayerNorm = nn.LayerNorm(config.hidden_size, eps=|\Suppressnumber|
|\addred|         config.layer_norm_eps)|\Reactivatenumber|
|\addred| -   self.dropout = nn.Dropout(config.hidden_dropout_prob)
|\addgreen| +   from epoi import FusedDropoutAddLayerNorm
|\addgreen| +   self.dropout_add_layernorm = FusedDropoutAddLayerNorm(|\Suppressnumber|
|\addgreen|         config.hidden_size, config.hidden_dropout_prob, eps=
|\addgreen|         config.layer_norm_eps)|\Reactivatenumber|
 
   def forward(self, hidden_states: torch.Tensor, input_tensor: torch.Tensor):
     hidden_states = self.dense(hidden_states)
|\addred| -   hidden_states = self.dropout(hidden_states)
|\addred| -   hidden_states = self.LayerNorm(hidden_states+input_tensor)
|\addgreen| +   hidden_states = self.dropout_add_layernorm(hidden_states, |\Suppressnumber|
|\addgreen|         input_tensor)|\Reactivatenumber|
     return hidden_states
\end{lstlisting}

\caption{The code transformation performed by \Tool to fuse Dropout and LayerNorm layers using the EPOI kernel}
\label{fig:example}
\end{figure}}


This section provides a motivating example for \Tool. Suppose Alice is a machine learning practitioner in an IT company. Her current project involves developing and training deep learning BERT-based models. As their product becomes increasingly popular, the number of model inference requests grows rapidly, which in turn causes a significant slowdown. 
After profiling, Alice estimates that reducing even 10\% off BERT's inference time could reduce their AWS bill by thousands of dollars per month. Therefore, Alice wants to find a solution to speed up their DL models.

After searching for solutions on the Internet, Alice finds several DL kernels that optimize Transformer-based DL models. Efficient PyTorch Operator Inventory (EPOI)~\cite{epoi} is one of them. It provides optimized replacements for PyTorch building blocks. For instance, it provides a kernel \CodeIn{FusedDropoutAddLayerNorm} to fuse \CodeIn{torch.nn.Dropout} and \CodeIn{torch.nn.LayerNorm} layers in a DL model. 
By fusing the operations, EPOI avoids redundant memory allocations and repeated tensor computation. It is reported that this kernel can accelerate model training by 12.2\%~\cite{epoi}. Alice is impressed with this runtime performance improvement and wants to update her BERT implementation with the fused kernel.

To apply this optimized kernel, Alice needs to manually search for the dropout and layernorm layers in her models and make the code changes. As she is maintaining multiple BERT-based models that can potentially apply this kernel, it is time-consuming to repeat the manual inspection. In each of her models, she needs to search for \CodeIn{torch.nn.Dropout} and \CodeIn{torch.nn.LayerNorm} within the code, determines whether they are used in tandem in the model structure, and edits the implementation. 

Alice tries to use the built-in search feature in her IDE. However, she finds that it cannot accurately match the consecutive ordering of two DL layers while tolerating naming variations and extraneous statements (e.g., \CodeIn{print} statements) that are not relevant to the model structure. As Alice edits the code to use the optimized kernels, she needs to carefully map the variables from the original code to the corresponding parameters of the optimized kernels. Such manual changes are tedious and error-prone, so she wants to find an automated approach to help her.

Alice turns to \Tool, a tool designed to automatically optimize model implementations using a diverse set of DL kernels including EPOI. She simply provides \Tool with her models' source code. \Tool scans through the codebase and identifies all locations where the fused module can be deployed to substitute the consecutive \CodeIn{Dropout} and \CodeIn{LayerNorm} layers. 
Within seconds, \Tool helps Alice find all the locations where consecutive \CodeIn{Dropout} and \CodeIn{LayerNorm} layers take place, as well as transforming these layers into the optimized kernel without syntax and semantic error. Furthermore, \Tool automatically detects and installs the missing packages and synthesizes import statements to the model source code. Alice reviews all changed locations and renames some variables.  Alice retrains all optimized models and gauges the computation efficiency improvement. Alice is happy with this automation, since all she needs to do is to review the optimization edits and check if they work. If not, she can simply revert the edits. This significantly simplifies her job and she no longer needs to deal with the tedious and error-prone manual optimization process anymore. 

\section{Approach}
\label{sec:Methodology}
\newcommand{\monosmall}[1]{\ttfamily{\footnotesize #1}}
\lstdefinestyle{dsl}{
  basicstyle=\itshape\small,
  columns=fullflexible,
  numbers=none,
  xleftmargin=1em,
  xrightmargin=0em,
  frame=none, 
  framerule=0.2pt,
  breakindent=1.5em,
  breaklines=true,
  literate={ε}{$\varepsilon$}{1}
           {|}{\ \ $|$\ \ }{3}
           {,}{ $,$ }{3}
           {.}{ $.$ }{3}
           {\$}{$\$$}{1}
           {(}{$($}{3}
           {[}{$[$}{3}
           {]}{$]$ }{3}
           {]]}{$)$}{3}
           {;}{ $;$ }{3}
           {=}{\monosmall{=}}{1}
           {+}{\monosmall{+}}{1}
           {-}{\monosmall{-}}{1}
           {@}{\monosmall{@}}{1}
           {prod}{$\cdot$}{1}
           {*}{\monosmall{*}}{1}
           {wildcard}{\monosmall{$\ast$}}{1}
           {...}{\monosmall{...}}{3}
           {string}{\monosmall{string}}{1}
           {number}{\monosmall{number}}{1}
           {identifier}{\monosmall{identifier}}{1}
           {<wild_card>}{\monosmall{<wild\_card>}}{1}
           {->}{\ \ $\rightarrow$\ \ }{4}
           {=>}{\ \ $\Rightarrow$\ \ }{4}
           {::=}{\ \ $::=$\ \ }{5}
           {\$sp\$}{ }{1}
}

\Tool includes three steps to automatically optimize the source code of a given model using DL kernels. Given a model implementation, \Tool first identifies the model structure by performing inter-procedural program analysis and represents the model structure in a DSL. Second, by leveraging a new algorithm powered by sub-sequence matching and scope analysis, \Tool identifies locations where a DL kernel is applicable. 
Finally, \Tool utilizes a synthesis-based code transformation approach to integrate the DL kernel into the model implementation.

\subsection{Domain-specific Language}
\label{sec:dsl}
\lstset{style=dsl}
\begin{figure}[t]
\begin{lstlisting}
Model_Structure::= DL_Node | DL_Node ; Model_Structure
DL_Node ::= DL_Layer | Tensor_Op 
DL_Layer ::= layer_type (input_list]]  -> variable
Tensor_Op ::= input op input -> variable | op input -> variable
layer_type ::= identifier | identifier . layer_type
input_list ::= input | input , input_list | ε 
input ::= argument | identifier = argument 
argument ::= variable | constant
op ::= + | - | @ | prod | ...
variable ::= identifier | identifier . variable
constant ::= string | number | ...
\end{lstlisting}
\caption{Domain-Specific Language}
\label{fig:dsl}
\end{figure}

\noindent Since model implementations are full of intricate syntax details and extraneous statements, it is challenging to directly match potential optimization locations to apply a kernel. Prior work~\cite{raghothaman2016swim, zhang2018code} has abstracted general-purpose programs as structured API call sequences through Domain-Specific Languages (DSL). Inspired by prior work, we design a DSL that abstracts program source code to sequences of DL layers and tensor operations to represent the model structure. Figure~\ref{fig:dsl} shows the DSL. We introduce the key terms in the DSL below. 


\lstdefinestyle{dsl_exp}{
  basewidth=0.53em,
  basicstyle=\ttfamily\footnotesize,
  columns=fullflexible,
  numbers=none,
  numbersep=1em,
  xleftmargin=2em,
  xrightmargin=0.5em,
  framexleftmargin=0.5em,
  framexrightmargin=0em,
  frame=single, 
  framerule=0.2pt,
  breakindent=2em,
  breaklines=true,
  rulesepcolor=\color{gray},
  literate={ε}{$\varepsilon$}{1}
           {,}{$,$}{1}
           {\$}{$\$$}{1}
           {(}{\thinspace $($}{1}
           {]]}{$)$}{1}
           {;}{$;$}{1}
           {=}{\ttfamily{=}}{1}
           {+}{\ttfamily{ + }}{1}
           {-}{\ttfamily{- }}{1}
           {@}{\ttfamily{@ }}{1}
           {*}{\ttfamily{*}}{1}
           {...}{\ttfamily{...}}{3}
           {[a-zA-Z][a-zA-Z0-9_]*}{\ttfamily{[a-zA-Z][a-zA-Z0-9\_]*}}{1}
           {string}{\ttfamily{string}}{1}
           {number}{\ttfamily{number}}{1}
           {->}{$\rightarrow$\ }{2}
           {<->}{\ $\leftrightarrow$\ }{2}
           {=>}{\ $\Rightarrow$\ }{2}
           {::=}{\ \ $::=$\ \ }{5}
           {torch.nn.Conv2d}{\ttfamily{torch.nn.Conv2d}}{1}
           {torch.nn.LayerNorm}{\ttfamily{torch.nn.LayerNorm}}{1}
           {torch.nn.Dropout}{\ttfamily{torch.nn.Dropout}}{1}
           {epoi.FusedDropoutAddLayerNorm}{\ttfamily{epoi.FusedDropoutAddLayerNorm}}{1}
           {self.dropout\_add\_layernorm}{\ttfamily{self.dropout\_add\_layernorm}}{1}
           {torch.nn.Linear}{\ttfamily{torch.nn.Linear}}{1}
           {torch.nn.BatchNorm1d}{\ttfamily{torch.nn.BatchNorm1d}}{1}
           {torch.nn.BatchNorm2d}{\ttfamily{torch.nn.BatchNorm2d}}{1}
           {torch.nn.intrinsic.qat.modules.LinearBn1d}{\ttfamily{torch.nn.intrinsic.qat.modules.LinearBn1d}}{1}
           {self.linear\_bn}{\ttfamily{self.linear\_bn}}{1}
           {slapo.op.linear.FusedQKV}{\ttfamily{slapo.op.linear.FusedQKV}}{1}
           {self.num\_heads}{\ttfamily{self.num\_heads}}{1}
           {self.num\_heads}{\ttfamily{self.num\_heads}}{1}
           {self.dense}{\ttfamily{self.dense}}{1}
           {self.LayerNorm}{\ttfamily{self.LayerNorm}}{1}
           {self.dropout}{\ttfamily{self.dropout}}{1}
           {self.dropout\_add\_layernorm}{\ttfamily{self.dropout\_add\_layernorm}}{1}
           {epoi.FusedConv2dBatchNorm2d}{\ttfamily{epoi.FusedConv2dBatchNorm2d}}{1}
           {self.conv\_batchnorm}{\ttfamily{self.conv\_batchnorm}}{1}
           {\$sp\$}{ }{1}
           {v_p}{$v_p$}{1}
           {v_s}{$v_s$}{1}
           {V}{$V$}{1}
           {A}{$A$}{1}
           {M}{$M$}{1}
           {R_i}{$\mathcal{R}_i$}{1}
           {R_c}{$\mathcal{R}_c$}{1}
           {:}{\ttfamily{:}}{2}
           {\{}{\ttfamily{\{}}{2}
           {\}}{\ttfamily{\}}}{2}
           {hidden\_states}{\ttfamily{hidden\_states}}{1}
           {input\_tensor}{\ttfamily{input\_tensor}}{1}
           {eps}{\ttfamily{eps}}{1}
           {kernel\_size}{\ttfamily{kernel\_size}}{1}
           {stride}{\ttfamily{stride}}{1}
}

\textbf{Model Structure $\mathcal{S}$.} A model structure is defined as a sequence of \textit{DL Nodes}. 
Below is an example model structure that invokes a \CodeIn{torch.nn.Linear} layer followed by a \CodeIn{torch.nn.Dropout} layer. The output tensor of the \CodeIn{Dropout} layer is added with another input tensor, which is then fed into a \CodeIn{torch.nn.LayerNorm} layer.  

\lstset{style=dsl_exp}
\begin{lstlisting}[firstnumber=1]
torch.nn.Linear(hidden_states]] $sp$ -> hidden_states;
torch.nn.Dropout(hidden_states]] $sp$ -> hidden_states; 
hidden_states + input_tensor $sp$ -> temp_1; 
torch.nn.LayerNorm(temp1]] -> hidden_states
\end{lstlisting}

\textbf{DL Node $\mathcal{N}$.} We use \textit{DL Node} as a general term to represent two types of computation units in a model structure---encapsulated DL layers (e.g., \CodeIn{torch.nn.Linear}, \CodeIn{torch.nn.Conv2d}) and primitive tensor operations (e.g., \CodeIn{+}, \CodeIn{-}, \CodeIn{@}). 

\textbf{DL Layer} ($\mathcal{L} (\{\alpha_i\}) \rightarrow \delta$). A DL layer is defined with the \textit{layer type} $\mathcal{L}$, as well as its input arguments $\{\alpha_i\}$ and an output variable $\delta$. We introduce \textit{variables} to normalize the input and output. Each input argument can be a \textbf{variable} or a \textbf{constant}. The output is a \textbf{variable}. We choose to preserve the original variable names and constant values in the DSL representation so that these concrete names and values can be later bound with abstract variables in a transformation rule to synthesize the transformed code. We use the fully qualified API name of a DL layer as the identifier of the layer. In the example above, since the model is implemented using PyTorch, we use the PyTorch API names to represent the layers. This design decision is made to distinguish different implementations of the same layer in a DL framework. For instance, PyTorch has two different implementations of the 2D convolution layers, \CodeIn{torch.nn.Cov2d} and \CodeIn{torch.nn.functional.conv2d}. 

\textbf{Tensor Op} ($ \alpha_1\ \mathcal{O}\ \alpha_2 \rightarrow \delta$ $ |\  \mathcal{O}\ \alpha_1 \rightarrow \delta$) is defined to represent primitive tensor operations, including tensor add (\CodeIn{+}), tensor multiplication (\CodeIn{@}), etc. A tensor operator can be a unary operator or a binary operator.  

One may wonder why we need to build such a DSL given that modern DL frameworks use intermediate representation (IR) to perform optimization. First, DL frameworks often lag behind the latest kernel developments. Many developers have started using kernels directly in their source code once these kernels were made available. Second, IR-level optimizations may not be stable and may even introduce errors, e.g., optimizing an inapplicable location due to incorrect pattern matching. Third, the DSL provides a high-level abstraction for model structures and eliminates many syntactic details that are unnecessary to match while identifying optimizable locations. By contrast, IRs are typically very detailed computational graphs where DL layers are broken down into atomic operators.

\subsection{Model Structure Abstraction}
\label{sec:abstraction}
Based on the DSL, we propose an inter-procedural analysis method to abstract DL model architectures from source code. In DL frameworks such as PyTorch and Keras, DL layers are initialized in an \CodeIn{init} function and then connected to compose a network in a composition function (e.g.,  the \CodeIn{forward} function in PyTorch and the \CodeIn{call} function in Keras). 
Therefore, our analysis method needs to accommodate this design principle when analyzing model implementations. We will use PyTorch as an example to explain our method in this section. 

\vspace{0.5em}
\noindent\textbf{Inter-Procedural Control Flow Analysis.} 
Given a model implementation, \Tool first performs inter-procedural control flow analysis to identify all control flow paths of each method in each class. Specifically, \Tool uses Google's Python Graphs ~\cite{python_graphs} to construct the control flow graph (CFG) of each method and connects them together based on method call relationships. \Tool then applies a depth-first search to identify all the control flow paths. To avoid infinite control flow paths caused by loops or recursions, \Tool stops its search in one direction if it encounters a control flow node that it has already visited. This approach ensures that loops are expanded only once in any given control flow path. 

By applying inter-procedural analysis, \Tool can handle cases where parts of a model structure are defined in a user-defined function that is called in the \CodeIn{forward} function. This is common in real-world model implementations, especially for models with complex structures. 
For each statement in a control flow path, \Tool filters it out if it is not a DL layer or a tensor operation. We manually went through the API documentation of PyTorch and identified all APIs that define a DL layer. This list of APIs is supplemented to \Tool to check whether a statement contains an API call that defines a DL layer. In this way, \Tool filters extraneous statements such as \CodeIn{print} and \CodeIn{log} statements. 

We observed that developers often define multiple model structures in one implementation and use configurations or command line arguments to control which structure to use at run time. This practice reduces the implementation effort and enhances the flexibility of their model structures since different models often share common building blocks. However, this makes the model implementation hard to analyze, since we need to identify all possible model structures from one implementation. For example, the \CodeIn{DPT} model from Hugging Face introduces a command line argument, \CodeIn{use\_batch\_norm}, to determine whether the model structure should include \CodeIn{BatchNorm} layers (Lines 5-7, 11-12, and 15-16 in Figure ~\ref{fig:example-structure-2}). When \CodeIn{self.use\_batch\_norm} is set to \CodeIn{true}, two \CodeIn{torch.nn.BatchNorm2d} layers are initialized in \CodeIn{init} and added into the model structure in \CodeIn{forward}. Otherwise, these two \CodeIn{BatchNorm} layers are excluded from the model structure. 

\Tool handles such cases by attracting model structures from all possible control paths that define a DL layer or a tensor operation. In  
Figure ~\ref{fig:example-structure-2}, since there are two possible control paths in the \CodeIn{forward} function, two possible model structures are extracted from the \CodeIn{forward} function. 

{\begin{figure}[t]
\vspace{4pt}
\footnotesize
\lstset{style=code}
\begin{lstlisting}[language=Python, firstnumber=1, escapechar=|]
 class DPTPreActResidualLayer(nn.Module):
   def __init__(self, config):
      self.convolution1 = nn.Conv2d(config.hidden_size, config.hidden_size, kernel_size=3)
      self.convolution2 = nn.Conv2d(config.hidden_size, config.hidden_size, kernel_size=3)
      if self.use_batch_norm:
         self.batch_norm1 = nn.BatchNorm2d(config.hidden_size)
         self.batch_norm2 = nn.BatchNorm2d(config.hidden_size)
\end{lstlisting}
\lstset{style=dsl_exp}
\begin{lstlisting}[firstnumber=1]
torch.nn.Conv2d(config.hidden_size, config.hidden_size, 3]] 
-> self.convolution1;
torch.nn.Conv2d(config.hidden_size, config.hidden_size, 3]] 
-> self.convolution2
\end{lstlisting}
\begin{lstlisting}[firstnumber=1]
torch.nn.Conv2d(config.hidden_size, config.hidden_size, 3]] 
-> self.convolution1;
torch.nn.Conv2d(config.hidden_size, config.hidden_size, 3]] 
-> self.convolution2;
torch.nn.BatchNorm2d(config.hidden_size]] -> self.batch_norm1;
torch.nn.BatchNorm2d(config.hidden_size]] -> self.batch_norm2;
\end{lstlisting}
(a)
\lstset{style=code}
\begin{lstlisting}[language=Python, firstnumber=8, escapechar=|]
   def forward(self, hidden_state: torch.Tensor) -> torch.Tensor:
      ...
      hidden_state = self.convolution1(hidden_state)
      if self.use_batch_norm:
         hidden_state = self.batch_norm1(hidden_state)
      ...
      hidden_state = self.convolution2(hidden_state)
      if self.use_batch_norm:
         hidden_state = self.batch_norm2(hidden_state)
\end{lstlisting}
\lstset{style=dsl_exp}
\begin{lstlisting}[firstnumber=1]
torch.nn.Conv2d(hidden_state]] -> hidden_state;
torch.nn.Conv2d(hidden_state]] -> hidden_state
\end{lstlisting}
\begin{lstlisting}[firstnumber=1]
torch.nn.Conv2d(hidden_state]] -> hidden_state;
torch.nn.BatchNorm2d(hidden_state]] -> hidden_state;
torch.nn.Conv2d(hidden_state]] -> hidden_state;
torch.nn.BatchNorm2d(hidden_state]] -> hidden_state
\end{lstlisting}
(b)
\caption{Code snippets from Hugging Face \texttt{DPT} model and the two possible model structures extracted from it.}
\label{fig:example-structure-2}
\end{figure}}

\vspace{0.5em}
\noindent\textbf{Alias Analysis and Type Resolution.} During the control flow analysis, \Tool keeps track of all defined variables, especially those defined in the \CodeIn{init} function. It also checks assignment statements to keep track of aliases. In other words, if the value of one variable is directly assigned to another variable without other computation, these two variables are considered as aliases. Keeping track of the definitions and aliases helps \Tool resolve the real value of variables when extracting model structures. 

For instance, when visiting the \CodeIn{init} function in Figure~\ref{fig:example-structure-2}, \Tool binds the initialized \CodeIn{torch.nn.Conv2d} object to \CodeIn{self.convolution1} (Line 3). Later, when visiting Line 10 in the \CodeIn{forward} function, \Tool resolves the function call on \CodeIn{self.convolution1} to \CodeIn{torch.nn.Conv2d}. Without this analysis, \Tool would not be able to resolve the actual type of \CodeIn{self.convolution1} and accurately extract the model structure. 

\vspace{0.5em}

\noindent\textbf{Handling nested method calls or tensor operations.}
When implementing a DL model, developers may write nested API calls to connect two layers or embed a tensor operation as an input argument to a DL layer. Below is an example to illustrate this:

\lstset{style=code}
\begin{lstlisting}[language=Python, numbers=left, firstnumber=5, belowskip=-5pt, escapechar=|]
   self.dropout = nn.Dropout(config.hidden_dropout_prob)
   self.LayerNorm = nn.LayerNorm(config.hidden_size, eps=config.layer_norm_eps)|\Suppressnumber|
   ...|\Reactivatenumber|
\end{lstlisting}
\begin{lstlisting}[language=Python, numbers=left, firstnumber=15, escapechar=|]
   hidden_states = self.LayerNorm(self.dropout(hidden_states) + input_tensor)
\end{lstlisting}

In this example, Line 15 first applies a \CodeIn{torch.nn.Dropout} layer on the hidden states and then adds the dropout output with the input tensor. The addition result is then passed into a \CodeIn{torch.nn.LayerNorm} layer. To handle such cases, \Tool performs depth-first traversal on the abstract syntax tree of the nested API call expression to unchain the nested API calls and tensor operations while introducing temporary variables to store the intermediate outputs. Specifically, it first extracts the most inner API call to \CodeIn{torch.nn.Dropout} as a DL layer and stores its output to a temporary variable \CodeIn{temp\_1}. Then, it extracts the \CodeIn{Tensor Add} operation into a tensor operation node with an intermediate output \CodeIn{temp\_2}. Finally, it extracts \CodeIn{torch.nn.LayerNorm} as a DL layer. The extracted model structure is listed as follows:

\lstset{style=dsl_exp}
\begin{lstlisting}[numbers=none]
torch.nn.Dropout(hidden_states]] $sp$ -> temp_1; 
hidden_states + input_tensor $sp$ -> temp_2; 
torch.nn.LayerNorm(temp2]] -> hidden_states
\end{lstlisting}

\subsection{Model Pattern and Transformation Rule}
This section describes how we can use the DSL to specify the optimization rules to integrate DL kernels into a model implementation. We first introduce the definition of \textbf{Pattern} and \textbf{Model Transformation Rule}. 

\label{sec:matcing}


We define the term \textbf{Pattern $\mathcal{P}$} to match with the model structures extracted from the source code of a model implementation. 
Similar to \textit{Model Structure} $\mathcal{S}$ (Section ~\ref{sec:dsl}), a \textbf{Pattern $\mathcal{P}$} consists of a sequence of DL nodes. To support fuzzy matching, we introduce wildcard ($\ast$) in a \textit{Pattern} to match with an arbitrary \textit{DL Node} in a model structure. This feature is useful when we define a pattern that does not care about certain nodes within a model structure. Furthermore, we use abstract variables starting with a dollar sign (\$) to represent input arguments and output variables in a a pattern. An abstract variable can be bound with a concrete variable name or a constant value during pattern matching. 

In DL frameworks such as PyTorch and Keras, DL layers are first initialized in an \CodeIn{init} function. Then, these layers are connected together in a composition function, e.g., the \CodeIn{forward} function in PyTorch and the \CodeIn{call} function in Keras. Therefore, when integrating a DL kernel into a model implementation, we need to transform both the \CodeIn{init} function and the composition function. Therefore, we define a \textbf{Model Transformation Rule} $\mathcal{TR}$ as a pair of rules $( \mathcal{R}_{i} , \mathcal{R}_{c} )$---$\mathcal{R}_{i}$ for transforming the \CodeIn{init} function and $\mathcal{R}_{c}$ for transforming the composition function. Each rule is further defined as a pair of model structure patterns $\mathcal{P}_{l} \Rightarrow \mathcal{P}_{r}$, where $\mathcal{P}_{l}$ denotes the model structure pattern before transformation (i.e., \textit{source pattern}) and $\mathcal{P}_{r}$ denotes the model structure pattern after transformation (i.e., \textit{target pattern}). 

Take the code transformation in Figure~\ref{fig:example}  as an example. The code changes in Figure~\ref{fig:example}  can be expressed as follows in our DSL:

\lstset{style=dsl_exp}
\begin{lstlisting}[firstnumber=1]
R_i: torch.nn.LayerNorm($s, $e]] -> $l; 
    torch.nn.Dropout($p]] -> $d 
    => 
    epoi.FusedDropoutAddLayerNorm($s, $p, $e]] -> $f
     
R_c: torch.nn.Dropout($a]] -> $x; 
    $x + $i -> $b; 
    torch.nn.LayerNorm($b]] -> $c 
    => 
    epoi.FusedDropoutAddLayerNorm($a + $i]] -> $a
\end{lstlisting}

The first rule $\mathcal{R}_{i}$ represents the transformation that substitutes the initialization of a \CodeIn{torch.nn.Dropout} layer and a \CodeIn{torch.nn.Layer-} \CodeIn{Norm} layer with the initialization of a single \CodeIn{FusedDropoutAddLayer-} \CodeIn{Norm} layer from the EPOI kernel.

The second rule $\mathcal{R}_{c}$ denotes that the structure of a \CodeIn{torch.nn.} \CodeIn{LayerNorm} layer followed by a tensor addition and a \CodeIn{Dropout} layer can be transformed into a \CodeIn{FusedDropoutAddLayerNorm} layer from the EPOI kernel. 
The transformation is considered complete only when both \textit{Rule Items} are satisfied and applied in the source code of a DL model.

Note that it is critical to keep all the abstracted variables in the patterns and transformation rules, since they denote the data flow between layers. They can also be used to prevent matching DL layers that match the layer types specified in a pattern but are not connected.

\subsection{Pattern Matching}

Given model structures abstracted from the source code of a DL model and a set of model transformation rules defined in \Tool,  
\Tool matches each model structure $\mathcal{S}$ with the source pattern of each transformation rule $\mathcal{TR}$. If a source pattern is matched, then it indicates the transformation rule is applicable to integrate the corresponding DL kernel into the model implementation. 

Since  $\mathcal{TR}$ has two parts, \Tool starts with matching $\mathcal{R}_{c}$, the rule for the composition function. \Tool starts with $\mathcal{R}_{c}$ instead of $\mathcal{R}_{i}$, since $\mathcal{R}_{c}$ specifies the ordering of DL nodes and the data flow between DL nodes, which prescribes more constraints for matching and thus helps us quickly narrow down irrelevant locations. By contrast, $\mathcal{R}_{i}$ only specifies the patterns in the \CodeIn{init} function, where the ordering of initialization statements does not matter and thus may lead to many false positive matches in the first place.

\Tool performs a subsequence matching between the source pattern of $\mathcal{R}_{c}$ and the model structures extracted from all composition functions (e.g., the \CodeIn{forward} function in PyTorch) in the model implementation. To match a DL node from the source pattern (denoted as $\mathcal{N}_{p}$) with a DL node (denoted as $\mathcal{N}_{s}$) in a model structure, \Tool first checks whether they have the same DL layer type or tensor operation type. If the type matches, \Tool then matches the input argument(s) and the output variable between $\mathcal{N}_{p}$ and $\mathcal{N}_{s}$. Note that the input argument(s) and output variable in $\mathcal{N}_{p}$ are abstract variables, while the input argument(s) and output variable in $\mathcal{N}_{p}$ are concrete names and variables. Thus, \Tool maintains a mapping $\mathcal{M}$ between the abstract variables and concrete names and values during matching. 

For each abstract variable in $\mathcal{N}_{p}$, \Tool first checks whether this variable has been bound with any existing concrete name or value in $\mathcal{M}$, since this abstract variable may have been used or defined in a previous DL node in the pattern and has already been bounded. If the abstract variable has not been bounded yet, \Tool binds it with the concrete name or value at the same position in $\mathcal{N}_{s}$ and put the mapping to $\mathcal{M}$. If the abstract variable has been bounded, \Tool checks whether the concrete name or value at the same position in $\mathcal{N}_{s}$ is the same as the previous bounded value. If not the same, the match fails. By ensuring the same abstract variable is bounded to the same concrete name or value in all locations in the pattern, \Tool ensures the dataflow consistency between the pattern and the matched model structure.

After successfully matching a subsequence in $\mathcal{S}$ with the nodes in the source pattern of $\mathcal{S}$ with $\mathcal{R}_{c}$, the next step is to identify the corresponding \CodeIn{init} function that initializes the matched DL layers. Then, \Tool binds the abstract variables in the source pattern of the transformation rule for the  \CodeIn{init} function, $\mathcal{R}_{i}$, with the concrete names and values in the abstraction from the identified \CodeIn{init} function in the model implementation. The resulting mappings will be used to perform code transformation in the next step.



    

        
                

        

\subsection{Synthesis-based Code Transformation}

\begin{figure}[t]
    \centering
    \includegraphics[width=0.8\linewidth]{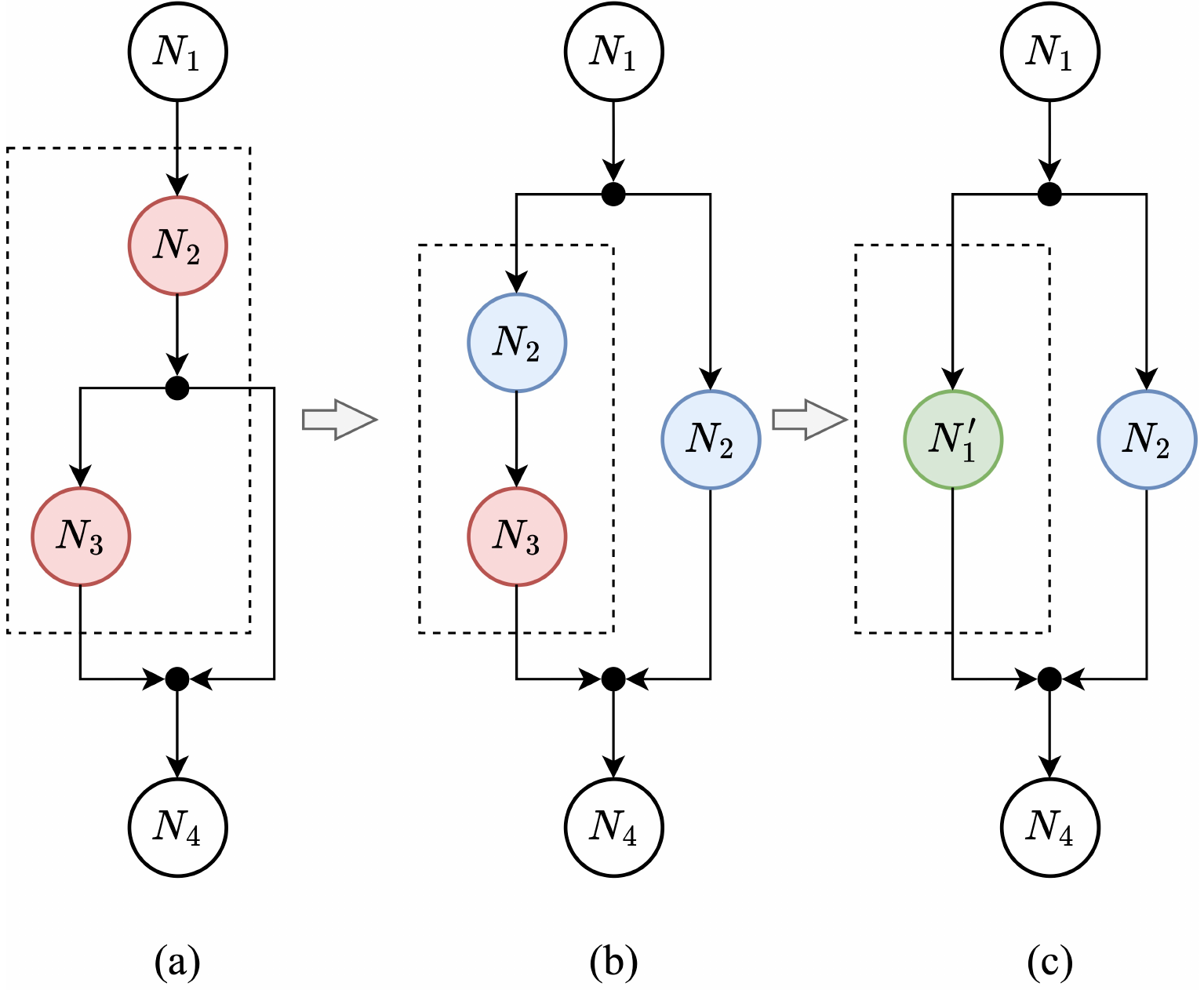}
    \caption{Replacing two statements in different basic blocks. (a) $\bm{N_3}$ lies in a branch while $\bm{N_2}$ does not; (b) \Tool makes two copies of $\bm{N_2}$ into the two branches; (c) \Tool replaces copied $\bm{N_2}$ and $\bm{N_3}$ with a new statement $\bm{N_{1}'}$}
    \label{fig:multi-path-transformation}
\end{figure}

\label{sec:refactor}
The objective of this step is to transform the original code in the matched locations into new code that makes use of the optimized DL kernel. Given the transformation rule of the DL kernel $\mathcal{TR}$, \Tool starts with transforming the \CodeIn{init} function, since it is relatively simple to transform and the optimized kernel needs to be initialized first before being used in the composition function. Based on the transformation rule for the \CodeIn{init} function, $\mathcal{R}_{i}$, \Tool first deletes the matched initialization statements in \CodeIn{init} and generates the initialization(s) for the optimized kernel. Specifically, for each abstract variable in the target pattern of $\mathcal{R}_{i}$, \Tool checks whether it is an abstract variable used in the source pattern. If it is, it means the variable is already defined in the model implementation. Then, \Tool looks up the mappings $\mathcal{M}$ generated from the previous pattern patching step to find out the concrete name or value of this variable and simply reuse it in the new code. If not, it means this is a new variable and \Tool synthesizes a variable name for this variable. 

After synthesizing the new initialization(s), \Tool moves on to transform the composition function based on $\mathcal{R}_{c}$. The transformation here is more complex, since the statements in the composition function are sensitive to the ordering. If the matched DL nodes in the composition function are next to each other and form a consecutive code block, \Tool can directly delete the corresponding statements and generate the new code with the same procedure as in the \CodeIn{init} function. However, in practice, the model implementations involve more complex structures that must be accounted for to ensure transformation correctness. 


\lstset{style=dsl_exp}
{\begin{figure}[tp]
\vspace{4pt}
\small
\begin{lstlisting}[firstnumber=1]
R_i: torch.nn.Conv2d($x]] -> $x; 
    torch.nn.BatchNorm2d($x]] -> $x 
    => 
    epoi.FusedConv2dBatchNorm2d($x]] -> $x

R_c: torch.nn.Conv2d($h, $h, kernel_size=$k]] -> $c;
    torch.nn.BatchNorm2d($h]] -> $bn 
    => 
    epoi.FusedConv2dBatchNorm2d($h, $h, kernel_size=$k]] -> $cb
\end{lstlisting}

(a) Model Transformation Rule $\mathcal{TR}$
\lstset{style=code}
\begin{lstlisting}[language=Python, firstnumber=1, escapechar=|]
 class DPTPreActResidualLayer(nn.Module):
   def __init__(self, config):
      self.convolution1 = nn.Conv2d(config.hidden_size, config.hidden_size, kernel_size=3)
|\addred| -    self.convolution2 = nn.Conv2d(config.hidden_size, config.|\Suppressnumber|
|\addred|          hidden_size, kernel_size=3)|\Reactivatenumber|
      if self.use_batch_norm:
|\addgreen| +       from epoi import FusedConv2dBatchNorm2d
|\addgreen| +       self.conv_batchnorm = FusedConv2dBatchNorm2d(config.|\Suppressnumber|
|\addgreen|             hidden_size, config.hidden_size, kernel_size=3)|\Reactivatenumber|
         self.batch_norm1 = nn.BatchNorm2d(hidden_size)
|\addred| -       self.batch_norm2 = nn.BatchNorm2d(hidden_size)
|\addgreen| +    else:
|\addgreen| +       self.convolution2 = nn.Conv2d(config.hidden_size,|\Suppressnumber|
|\addgreen|             config.hidden_size, kernel_size=3)|\Reactivatenumber|

   def forward(self, hidden_state: torch.Tensor) -> torch.Tensor:
      ...
      hidden_state = self.convolution1(hidden_state)
      if self.use_batch_norm:
         hidden_state = self.batch_norm1(hidden_state)
      ...
|\addred| -    hidden_state = self.convolution2(hidden_state)
      if self.use_batch_norm:
|\addred| -       hidden_state = self.batch_norm2(hidden_state)
|\addgreen| +       hidden_state = self.conv_batchnorm(hidden_state)
|\addgreen| +    else:
|\addgreen| +       hidden_state = self.convolution2(hidden_state)
\end{lstlisting}
(b) The transformation synthesized by \Tool

\caption{Transforming model \texttt{DPT} using \Tool.}
\label{fig:example-multiple}
\end{figure}}

\begin{table*}[t]
  \caption{Model transformation rules used for evaluation.}
  \label{table:rules}
  \begin{tabular}{ccccc}
    \toprule
    Index & Name & Description & Source \\
    \midrule
 1 & \CodeIn{BertSelfAttention} & \CodeIn{BertSelfAttention} with xformers' attention ops & \CodeIn{xformers}~\cite{xsformer} \\
     2 & \CodeIn{T5Attention} & \CodeIn{T5Attention} with xformers' attention ops & \CodeIn{xformers}~\cite{xsformer} \\
     3 & \CodeIn{GPT2Attention} & \CodeIn{GPT2Attention} with xformers' attention ops  & \CodeIn{xformers}~\cite{xsformer} \\
     4 & \CodeIn{softmax} & A drop-in replacement to \CodeIn{torch.nn.softmax} & \CodeIn{triton}~\cite{noauthor_triton_2023} \\
     5 & \CodeIn{Dropout\_LayerNorm} & Fusing \CodeIn{Dropout} and \CodeIn{LayerNorm} & \CodeIn{epoi}~\cite{epoi} \\
     6 & \CodeIn{biased\_GeLU} & Fusing biased \CodeIn{Linear} and \CodeIn{GeLU} activation & \CodeIn{epoi}~\cite{epoi} \\
     7 & \CodeIn{Conv\_BatchNorm} & Fusing \CodeIn{Conv2d} and \CodeIn{BatchNorm2d}  & \CodeIn{PyTorch}~\cite{pytorch} \\
     8 & \CodeIn{Linear\_BatchNorm} & Fusing \CodeIn{Linear} and \CodeIn{BatchNorm1d} & \CodeIn{PyTorch}~\cite{pytorch} \\
    9 & \CodeIn{fused\_QKV} & Fusing three \CodeIn{Linear} layers as \CodeIn{q}, \CodeIn{k}, and \CodeIn{v} in encoder & \CodeIn{slapo}~\cite{noauthor_slapo_2023} \\
    \bottomrule
  \end{tabular}%
\end{table*}

Take the implementation of DPT in Figure~\ref{fig:example-structure-2} as an example. Given the model transformation rule in Figure~\ref{fig:example-multiple}(a), the \CodeIn{torch.nn.Conv2d} layer is matched with the first Conv2d layer in the \CodeIn{forward} function (Line 10 in Figure~\ref{fig:example-structure-2}), which is before the \CodeIn{if} statement at Line 11. However, the \CodeIn{torch.nn.BatchNorm2d} layer is matched with the batch norm layer at Line 12, which is within the \CodeIn{if} statement. Therefore, \Tool cannot simply substitute them with a single fused kernel.

To handle these special cases, \Tool refactors the original implementation by moving the matched statement outside an \CodeIn{if} statement to both branches of the \CodeIn{If} statement so that they can be replaced with other ones without affecting the correctness of the program, as illustrated in Figure ~\ref{fig:multi-path-transformation}. Specifically, \Tool takes as input a control flow graph, where each statement is represented by a node, along with a set of nodes $\{N_1, N_2, ..., N_m\}$, which forms a subset of a single control flow path. \Tool visits these nodes from the beginning. When $N_i$ lies in a branch that does not contain the previous nodes, \Tool copies the previous nodes to the beginning of both branches where $N_i$ lies and removes them from their original position. \Tool continues this process until it reaches to the first converging point in the control flow graph. Following this, \Tool proceeds with a similar process to examine these nodes in a reverse sequence. When $N_j$ lies in a branch that does not contain the subsequent nodes, \Tool copies the subsequent nodes to the end of both branches where $N_j$ lies, and removes them from their original position, until \Tool reaches the first diverging node. At this point, \Tool can safely delete the original code snippets and generate new ones based on the target pattern of $\mathcal{R}_{c}$. As an example, Figure ~\ref{fig:example-multiple}(b) demonstrates the code transformation on model \CodeIn{DPT} (Figure ~\ref{fig:example-structure-2}) following the model transformation rule in Figure ~\ref{fig:example-multiple}(a).


\subsection{Implementation}


We implemented \Tool in Python with 1900 lines of code. \Tool employs Google's Python Graphs ~\cite{python_graphs} as its external dependency library. Our source code and experiment scripts have been made publicly available on Zenodo~\cite{zenodo_artifact} and GitHub.\textsuperscript{\ref{note1}}

\section{Evaluation}
\label{sec:Evaluation}
\newcommand{\hf}[1]{{\mytexttt{\scriptsize #1}}}

To evaluate the effectiveness of our proposed tool, we conducted experiments to answer three research questions:

\begin{itemize}[leftmargin=10mm]
    \setlength\itemsep{0.5mm}
    \item[\textbf{RQ1}] How accurate is \Tool in targeting locations applying optimization rules to DL models compared to existing automatic code refactor techniques? 

    \item[\textbf{RQ2}] How precise is \Tool in refactoring model source code to apply optimization rules compared to existing automatic code refactor techniques?

    \item[\textbf{RQ3}] To what extent can control-flow analysis and scope analysis affect the effectiveness of code pattern detection?

    \item[\textbf{RQ4}] How does the optimization rules used by \Tool affect the performance of DL models?

    \item[\textbf{RQ5}] How much runtime overhead does \Tool introduce?
\end{itemize}

\subsection{Benchmark}

\lstset{style=dsl,
  xleftmargin=0,
  frame=none, 
  backgroundcolor=\color{backcolour}
  }

\begin{table*}[t]
  \caption{Total number of optimizations applied on a set of 199 models in Hugging Face using 9 transformation rules with \Tool and PyEvolve. }
  \label{table:result}
  {%
  \begin{tabularx}{0.7\textwidth}{c|c|ccccc|ccccc}
    \toprule
    Rule & Ground & \multicolumn{5}{c|}{\Tool} & \multicolumn{5}{c}{Baseline} \\
    Index & Truth & TP & FN & FP & Precision & Recall & TP & FN & FP & Precision & Recall \\
    \midrule
     1 & 1&   1&0&0    & 1&1       & 1&0&0     & 1&1   \\
     2 & 2&   2&0&0    & 1&1       & 2&0&0     & 1&1   \\
     3 & 2&   2&0&0    & 1&1       & 1&1&0     & 1&0.50   \\
     4 & 208& 197&11&0  & 1&0.95    & 137&71&0  & 1&0.66   \\
     5 & 80&  80&0&0   & 1&1       & 78&1&9    & 0.90&0.99    \\
     6 & 143& 136&7&0 & 1&0.95    & 66&77&0   & 1&0.46    \\
     7 & 35&  31&4&0   & 1&0.89    & 2&33&0    & 1&0.06  \\
     8 & 2&   2&0&0    & 1&1       & 0&2&0     & -&0.00   \\
     9 & 2&   2&0&0    & 1&1       & 2&0&1     & 0.67&1   \\
     \midrule
   Sum & 475& 453&22&0 & 1&0.95 & 289&185&10 & 0.97&0.61 \\
    \bottomrule
  \end{tabularx}%
  }
\end{table*}

To evaluate \Tool, we concluded a set of 9 optimization rules from open-source repositories~\cite{epoi,noauthor_triton_2023,xsformer} and academic papers~\cite{noauthor_slapo_2023}, as listed in Table ~\ref{table:rules}, and used 199 open-sourced DL models in Hugging Face (git version \CodeIn{ef42cb627}). 

\textbf{Optimization Rules.} xFormers~\cite{xsformer} provides drop-in replacements for basic DL operators, such as \CodeIn{softmax} and attention operator. Based on the flash attention operator provided by xFormers, epoi~\cite{epoi} implements and encapsulates replacement attention operators for transformer-based models, e.g. GPT2, Bert, and T5. Both epoi and slapo~\cite{noauthor_slapo_2023} provide fused operators that can accelerate model training by reducing throughput. 

\textbf{Models.} The GitHub repository~\cite{hf_transformers} of Hugging Face transformers contains 224 DL models. We omitted 25 models from our analysis due to the absence of their source code implementation.

\textbf{Dataset.} Among the remaining 199 Hugging Face models with source code, two of the authors independently spent approximately 20 hours each to identify locations in the source code where the 9 optimization rules are applicable. They achieved a high percentage of agreement of 96.9\%. Following this, the authors collaboratively reviewed any discrepancies, resulting in a refined dataset of 463 locations across 199 DL models where the 9 optimization rules can be applied. This dataset serves as the basis for assessing the performance of \Tool.

\subsection{Environmental Setups}
We performed our evaluation on an Amazon EC2 instance with an Intel(R) Xeon(R) CPU and a Tesla T4 GPU. The operating system is 64-bit Ubuntu 20.04 LTS with 
Python 3.8.10 and CUDA 11.7. Our Python virtual environment is installed with PyTorch 2.0.1 and the latest version of apex~\cite{apex}, xFormers~\cite{xsformer}, epoi~\cite{epoi} and slapo~\cite{noauthor_slapo_2023}. We also installed the latest version of Comby~\cite{comby} as the dependent tool of our baseline approach.

\subsection{Comparison Baseline}
We use PyEvolve~\cite{dilhara_pyevolve_2023} as our comparison baseline. PyEvolve is a state-of-the-art approach for inferring Python code transformation rules based on example code changes and it uses Comby~\cite{comby} to automatically apply the rules to new code snippets. While their rule inference tool is not yet open-sourced, we manually created one code transformation rule for each optimized kernel to deploy it on one of the DL models. We then applied the rules to other models in our benchmark to evaluate the effectiveness of PyEvolve. 


\subsection{Results}

\begin{table*}[t]
  {%
  \caption{The total number of optimizations applied on 119 Hugging Face models using 9 transformation rules.*}
  \label{table:result-ablation}
  \begin{tabular}{c|cccccccccc|ccc}
    \toprule
    Rule Index & 1 & 2 & 3 & 4 & 5 & 6 & 7 & 8 & 9 & All & Precision & Recall \\
    \midrule
    \Tool & 1/0/0 & 2/0/0 & 2/0/0 & 197/11/0 & 80/0/0 & 136/7/0 & 31/4/0 & 2/0/0 & 2/0/0 & 453/22/0 & 1.00 & 0.95 \\
    \textsc{Adopter}$_P$ & 1/0/0 & 2/0/0 & 2/0/0 & 197/11/1 & 80/0/0 & 25/118/0 & 25/10/0 & 2/0/0 & 2/0/0 & 336/139/1 & 1.00 & 0.71 \\
    \bottomrule
  \end{tabular}%
  \vspace{2pt}
  
  {\small *The numbers are presented in the form of True Positive / False Negative / False Positive.}
  }
\end{table*}

\subsubsection{Effectiveness of Pattern Matching.}
\label{sec:effectiveness-matching}
We used both \Tool and PyEvolve, our baseline approach, to apply the 9 model optimization rules on 199 Huggingface models. We then compare the results with our dataset to compute the precision and recall of applying model optimization rules using these two tools. Specifically, a true positive indicates that a tool successfully applies an optimization rule on a model at the location identified in our dataset. 

The result is shown in Figure ~\ref{table:result}. The average recall of applying optimization rules across the 199 models with 9 optimization rules is 95.4\% using \Tool and 61.0\% using the baseline approach. The average precision is 100.0\% using \Tool and 96.7\% using the baseline approach. \Tool improves the precision and recall by 3\% and 56\%, respectively. 

The reason why \Tool significantly outperforms our baseline in terms of recall is two-fold. First, as the rules used by PyEvolve are generated from code change examples, it requires the function names of the method calls are identical to those in the example. To illustrate, the code transformation rule used by PyEvolve to apply the 4th optimization rule in Table~\ref{table:rules} can only identify the method calls whose function name is exactly \CodeIn{nn.functional.softmax}. Therefore, it fails to identify the method calls in other models which invoke \CodeIn{F.softmax}, \CodeIn{torch.nn.functional.} \CodeIn{softmax}, etc. This is because in Python, an external library is imported with an \CodeIn{import} statement, which enables developers to assign aliases to the packages to make them more convenient to use. In comparison, \Tool resolves the fully qualified name of a DL layer when abstracting model structures, which enables it to identify all method calls that use the identical DL layer. 

Second, the rules employed by PyEvolve are not flexible in identifying the same model structure with different code patterns. For example, the rule used by PyEvolve to apply the 7th optimization rule in Table ~\ref{table:rules} identifies two consecutive method calls of a \CodeIn{torch.nn.Conv2d} and \CodeIn{torch.nn.BatchNorm2d} layers. As PyEvolve matches and transforms source code based on the fixed code pattern, it fails to identify the potential code change discussed in Figure ~\ref{fig:example-multiple}, in which the method call of the \CodeIn{torch.nn.BatchNorm2d} layer lies in one branch of an \CodeIn{if} statement and \CodeIn{torch.nn.Conv2d} lies outside the \CodeIn{if} statement. On the contrary, \Tool can identify this situation, as it employs control-flow analysis to extract model structure and match patterns in every control flow path.

\begin{finding}{}
Compared to PyEvolve, the state-of-the-art example-based code refactoring technique, \Tool improves the recall by 56\% on our benchmark while improving the precision by 3\%.
\end{finding}

\subsubsection{Correctness of Code Transformation.}
\label{sec:eval-refactor}
To assess patch correctness, ideally, we can apply the patches to DL models, execute the models, and compare their behavior. However, it is time-prohibitive to repeat this process for all 199 models. This is because, for each model, we need to download the dataset for training, debug and fix issues in the training script, or even write new training scripts. Note that this is different from loading a pre-trained model, since model structures are changed after optimization and we need to build, run, and train the model from scratch. Therefore, we assess their correctness through manual inspection.


In particular, we manually inspected the 453 patches generated by \Tool and the 298 patches generated using the approach of PyEvolve. We first found that as PyEvolve uses Comby to match and apply code changes based on a fixed code pattern, the patches generated with our baseline approach are 100\% correct. To facilitate a consensus on the correctness of patches generated by \Tool, the two authors separately inspected the correctness of all the 453 patches and discussed where they disagreed. The agreement level between the two raters is 0.93 in terms of Cohen’s Kappa~\cite{cohenskappa}, indicating a substantial agreement level.  This manual process took about 4 person-hours. By leveraging the inter-rater agreement, we aim to minimize errors caused by overlooking details in such evaluative tasks. 

As a result, 13 of 453 patches are deemed incorrect. The correctness rate is 97.1\%. Though not 100\%, it is considerably high given that \Tool significantly improves the recall by 53\% while keeping the high precision (97\%) of the template-based code pattern matching approach. We further examined the incorrect patch. Surprisingly, we found that they are all caused by the functionality flaw of python-graphs, the control flow analysis tool we used in \Tool. While analyzing those model source codes, python-graphs failed to correctly keep the line number information, leading to the false localization in the incorrect patches.

\begin{finding}{}
Compared to the template-based code transformation approach used by PyEvolve, our synthesis-based approach correctly generates 97.1\% of the patches. 
\end{finding}

\subsubsection{Ablation Study}
We conducted an ablation study to evaluate the effectiveness of the inter-procedural control-flow analysis method (Section ~\ref{sec:abstraction}) used in \Tool. Specifically, we designed a variation of \Tool, \textsc{Adopter}$_P$, in which we replaced the control flow path analysis with the flattening control flow nodes in a list, while also eliminating the analysis on nested method calls. The result in Figure ~\ref{table:result-ablation} shows that while the precision of \textsc{Adopter}$_P$ does not decrease, its recall decreases from 95.4\% to 70.7\%. This result illustrates the significance of our inter-procedural control-flow analysis method. 

\begin{finding}{}
Our ablation study shows that the inter-procedural control-flow analysis leveraged by our approach can significantly improve the effectiveness of \Tool.
\end{finding}

\begin{figure*}[htbp]
    \centering
    \begin{subfigure}{0.35\textwidth}
        \centering
        \includegraphics[height=1.3in]{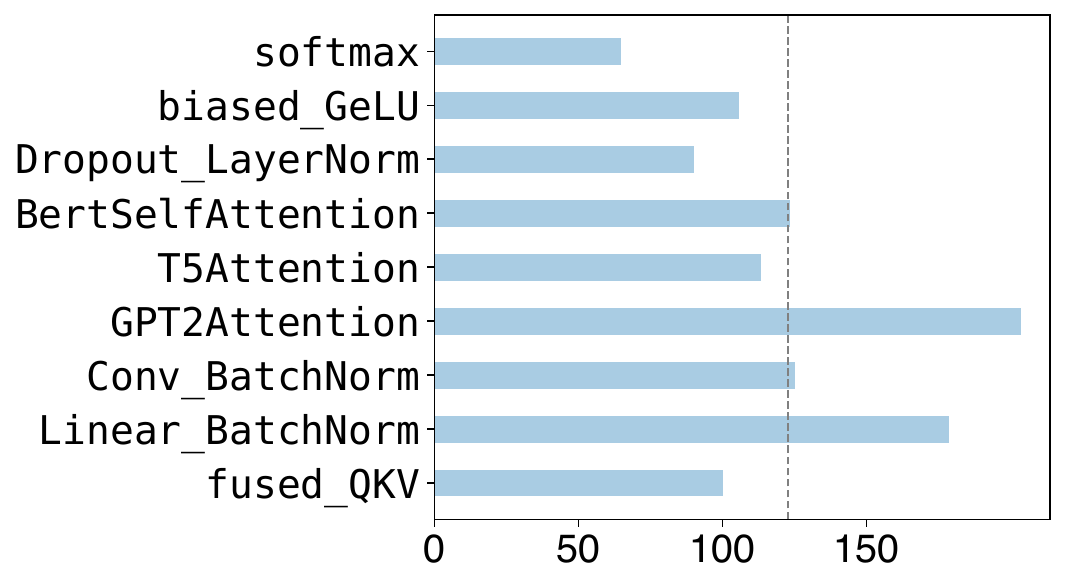}
        \caption{Sample per Second (\%)}
    \end{subfigure}%
    ~ 
    \begin{subfigure}{0.35\textwidth}
        \centering
        \includegraphics[height=1.3in]{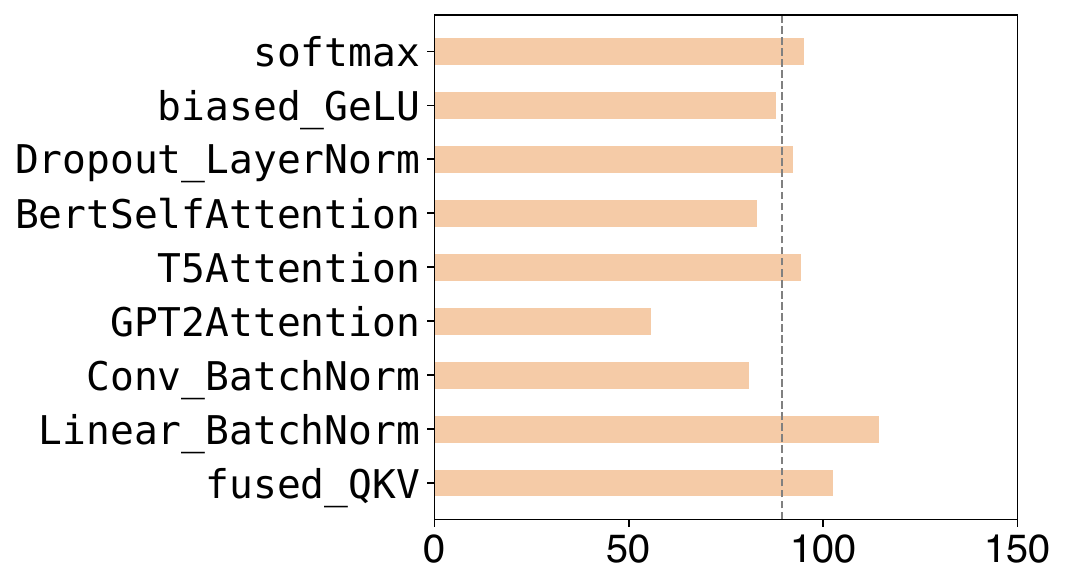}
        \caption{per GPU Memory (\%)}
    \end{subfigure}
    \begin{subfigure}{0.25\textwidth}
        \centering
        \includegraphics[height=1.3in]{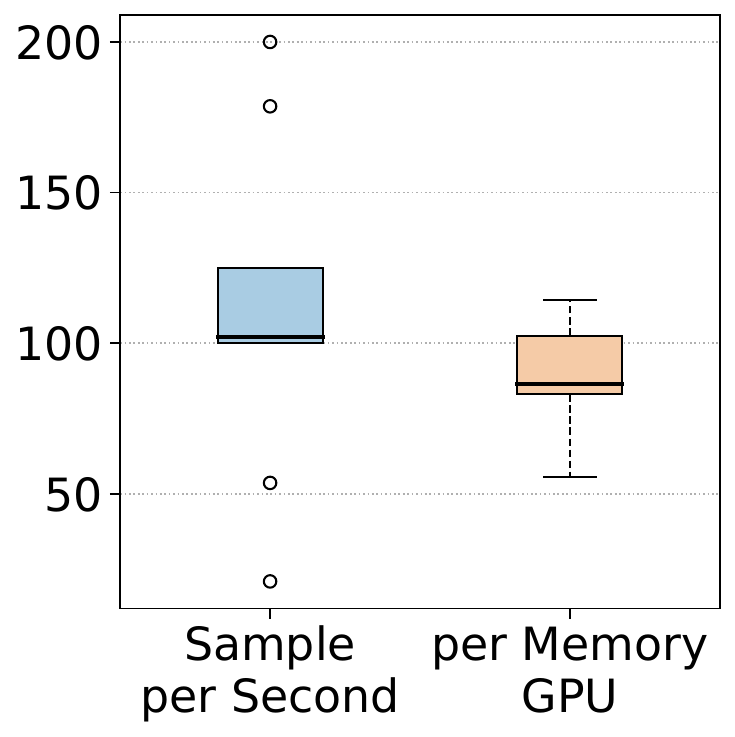}
        \caption{Model Performance (\%)}
    \end{subfigure}
    \caption{Effectiveness of optimization rules. (a)(b) Average change of model performance after applying each of the 9 optimization rules. The dashed line indicates the overall average change. (c) Distribution of changes in model performance after applying all optimization rules.}
    \label{fig:performance}
\end{figure*}

\subsubsection{Effectiveness of Optimization Rules.}
\label{sec:effectiveness}
We investigated the impact on the performance of nine Hugging Face models after applying the optimization rules to reflect their effectiveness. For each model, we trained one version applied with each optimization rule that is successfully matched using \Tool. Besides, we also trained a basic version without any optimization (referred to as the ``vanilla'' version), and an ``aggregate'' version where all relevant optimization rules were applied. We compared two metrics: 

\begin{enumerate}[leftmargin=0.6cm]
    \item[1)] {\em Sample per Second}, which measures the average amount of training data samples a DL model processes in each second. A higher value means faster training. 
    \item[2)] {\em Per GPU Memory}, which measures the average memory usage (in GB) of each GPU during training. A lower value means better memory efficiency.
\end{enumerate}

We calculate the increase in training speed by dividing the number of {\em Sample per Second} of the optimized model by that of the vanilla version. On the contrary, we calculate the increase in memory efficiency by dividing the number of {\em per GPU Memory} of the vanilla model by that of the optimized ones. For a more accurate assessment, we trained each model for 20 iterations and then calculated the evaluation metrics by averaging the results over 1 iteration. 


Figure ~\ref{fig:performance} demonstrates the average performance increase using each optimization rule. Most of the optimization rules can help increase the training speed or increase the memory efficiency. On average, these rules help increase the Sample per Second metric by 22.7\% and decrease the per GPU Memory metric by 10.5\%. 
The reasons why some of the optimization rules do not help with the model performance are two-fold. First, for the \texttt{fused\_QKV} kernel, the computation of each \texttt{q}, \texttt{k}, and \texttt{v} matrix multiplication is already saturated on a single GPU, so the advantage of \texttt{fused\_QKV} is mainly to reduce the communication overheads when training a large model with model parallelism, in which \texttt{q}, \texttt{k} and \texttt{v} are partitioned to multiple GPUs. Currently, we have not included model parallelism in \Tool, and we are leveraging this feature in future work. Second, with the continuous evolution of DL frameworks, there is an ongoing improvement in the efficiency of their built-in kernels. This enhancement could lead to these native kernels surpassing the performance of some currently optimized kernels.

\begin{finding}{}
On average, the optimizations applied with \Tool help improve the model performance by increasing the {\em Samples per Second} metric by 22.7\% and decreasing the {\em per GPU memory} metric by 10.5\%. 
\end{finding}

\subsubsection{Runtime Overhead}

\begin{table}[t]
  {%
  \caption{Statistics of average \Tool running time (in seconds) on 199 Hugging Face models with 9 optimization rules.}
  \label{table:result-overhead}
  \begin{tabular}{c|c|c|c|c}
    \toprule
    Maximum & Minimum & Mean & Median & Std. Dev \\
    \hline
    7.86 & 0.03 & 1.34 & 1.23 & 1.08 \\
    \bottomrule
  \end{tabular}%
  }
\end{table}

One may wonder how the runtime overhead of \Tool compares to the efficiency gains from optimizations, and whether it is practical for production use. To answer this question, we calculated the average running time (in seconds) of running \Tool on each of the 199 Hugging Face models with 9 optimization rules. Table \ref{table:result-overhead} shows the statistics of the results.  

Overall, \Tool is very fast. Given a Hugging Face model and an optimization rule, it takes an average of 1.34 seconds to perform source code analysis, identify a location to apply a kernel, and execute the code transformation. This runtime overhead is minimal compared with model training, which usually takes hours. As shown in Section \ref{sec:effectiveness}, \Tool can enhance training speed by 22.7\%. Thus, we believe \Tool is useful and also practical in production. 

\begin{finding}{}
On average, \Tool takes a few seconds to execute on a Hugging Face model with an optimization rule, which is trivial compared to the efficiency gains on model training.
\end{finding}

\section{Discussion}
\label{sec:Discussion}
\subsection{Tool Design for DL Optimization}
\Tool is designed to mitigate the challenge of manually searching and adopting optimized DL kernels. This challenge arises from several factors. Firstly, DL practitioners often favor optimization solutions that are easy to debug. Despite numerous optimization pipelines capable of accelerating models, users remain curious about the underlying mechanisms of these pipelines. Therefore, developers continue to produce refined, ready-to-use optimized kernels compatible with popular DL frameworks, such as PyTorch. Second, Secondly, the design of these optimized DL kernels is becoming increasingly intricate. As a consequence, most developers prefer to learn and apply these optimizations to their model source code through simple instructions and prompts. Last but not least, the implementation of a specific DL model architecture incorporates a wide variety of coding approaches and styles. This makes it difficult to apply changes from one model's patches to another. These insights indicate that automated DL optimization tools should be able to help users better understand the optimization solutions. Furthermore, these tools are expected to handle DL models implemented in various coding styles. Hugging Face is considered a high-quality benchmark for such tasks, given its extensive repository of open-source DL models developed by a global community of DL practitioners.

\subsection{Threats to Validity}
Regarding {\em external validity}, one threat is that we have only evaluated \Tool on nine kernels developed for PyTorch. While PyTorch is widely used in the DL community, many popular DL models are also developed on other platforms, e.g., Keras, TensorFlow, MXNet, etc. To enhance the generalizability of our findings, we will extend our evaluation to more kernels from other platforms in the future.


Regarding {\em construct validity}, a threat is that we only manually inspected the transformations generated by \Tool and the baseline approach to evaluate their correctness (Section \ref{sec:eval-refactor}). Ideally, an alternative evaluation method is to re-train the optimized models and compare their behavior with the original models before transformation. However, it is prohibitively time-consuming, since for each model, we need to download the dataset, debug the training script, or even write new scripts from scratch. 
To mitigate this threat, we applied the patches to nine popular Hugging Face models and successfully trained the models (Section \ref{sec:effectiveness}). This could serve as evidence of the correctness of these patches. This finding, to some extent, supports the validity of our manual inspection approach in assessing patch correctness.



\subsection{Limitations and Future Work}
\Tool has been specifically designed to implement model transformation rules that substitute DL layers and tensor operations. As outlined in Section \ref{sec:effectiveness}, a notable limitation of \Tool in its current form is its inability to handle tensor parallelism. We plan to extend our approach to support tensor parallelism in the future.

\Tool tracks equivalent conditions when extracting model structures. For example, in Figure \ref{fig:example-structure-2}, \Tool extracts two model structures because there are two \CodeIn{if} conditions with the same guard expression. The current implementation of \Tool checks whether the guard conditions are the same at the text level, since guard conditions in model implementation are often simple, e.g., checking a model configuration variable. This can be improved by checking the logic equivalence of two guard conditions using an SMT solver like Z3 \cite{z3paper}.


Finally, in this work, we manually specified the DL optimization rules, since we only needed to create only one rule per kernel and it’s a one-time effort. Theoretically, these optimization rules can also be automatically inferred by leveraging existing example-based inference approaches \cite{meng_systematic_2011, meng_lase_2013} with our DSL abstraction. This will be considered as part of our future work.



\section{Related Work}
\label{sec:RelatedWork}
\subsection{DL Optimization}

There has been a wide variety of Deep Learning model optimization solutions. Apex~\cite{apex} by Nvidia provides a library implemented at the bottom layer for mixed precision and distributed training. Proposed by Meta, xFormers~\cite{xsformer} provides solutions to optimize Transformer-based DL models with independent and customizable building blocks. Deepspeed ~\cite{aminabadi2022deepspeed} by Microsoft proposed and implemented ZeRO~\cite{rajbhandari2020zero}, a data parallelism approach to optimize memory to optimize memory, improve training speed, and increase the trainable model size. Megatrom-LM~\cite{shoeybi2020megatronlm} by Nvidia proposed an intra-layer model parallel approach for Transformer-based models. PyTorch proposed \CodeIn{torch.fx}~\cite{reed2022torchfx}, a self-integrated intermediate representation (IR) to help with optimization on the computation graph level. To help integrate various optimization solutions, Chen et al.~proposed a scheduling language called slapo~\cite{noauthor_slapo_2023}, which provides on-demand and progressive optimizations in the model execution phase.

\subsection{Automated Code Transformation}

The code transformation system in \Tool is related to existing automated code transformation techniques \cite{meng_systematic_2011, meng_lase_2013, zhang2017automated, rolim2017refazer, ketkar_inferring_2022, dilhara_pyevolve_2023} following the learning-from-example approach. Specifically, these approaches automatically generate code or infer code transformation rules from example code changes. 
Meng et al.~proposed \textsc{Sydit} \cite{meng_systematic_2011}, an automated approach that applies similar edits to similar locations based on a single example. \textsc{LASE} \cite{meng_lase_2013} extends \textsc{Sydit} by inferring an edit script from multiple examples. Rolim et al.~proposed \textsc{ReFazer}, which synthesizes program transformations from examples with a domain-specific language. 
\textsc{APIfix}~\cite{gao2021apifix} automates API usage adaptations by learning from example human adaptations from the old library version to the new library version, as well as example usages of newly updated libraries. 
Ketkar et al.~proposed \textsc{TC-Infer} \cite{ketkar2022inferring} to infer type change patterns from the version history of open-source projects.
Dilhara et al.~proposed \textsc{CPATMiner} ~\cite{pythoninfer} to automatically infer code change patterns in Python programs. They further developed \textsc{PyEvolve} \cite{dilhara_pyevolve_2023} to identify similar locations in new Python codebases and apply code change patterns.
While these techniques are effective in automating minor, small changes in both Java and Python code snippets, they are faced with flexibility issues when it comes to DL optimization. 
\Tool managed to address this challenge by abstracting the model source code into DL Model Structures.

\section{Conclusion}
\label{sec:Conclusion}
This paper proposed \Tool, an automated Deep Learning model optimization approach to help developers deploy optimized kernels on the source code level. We design a Domain-Specific Language (DSL) for representing DL model architectures and leverage this DSL to specify model transformation rules required to integrate a DL kernel into a DL model. \Tool leverages an inter-procedural code analysis to abstract the model structure using our DSL. Then, \Tool performs scope analysis and sub-sequence matching to identify locations in the model structure where the transformation rules can be applied. Finally, \Tool proposes a synthesis-based code transformation method to apply the transformation rule on the model source code. Our evaluation on a benchmark of 9 model optimization rules and 199 Hugging Face models shows that compared to a state-of-the-art automated code transformation technique, \Tool helps improve the precision and recall of applying optimization rules on the DL models by 3\% and 56\%, respectively. It also revealed that, on average, our optimization rules help improve the model performance by increasing the {\em Samples pre Second} metric by 22.7\% and decreasing the {\em per GPU Memory} metric by 10.5\%. 

\section{Data Availability Statement}
Our code and experiment data are made publicly available on Zenodo~\cite{zenodo_artifact} and also on GitHub: \href{https://github.com/ailen-wrx/Adopter}{https://github.com/ailen-wrx/Adopter}.

\begin{acks}
The authors would like to thank the reviewers for their valuable comments. This research was in part supported by an Amazon Research Award.
\end{acks}

\bibliographystyle{ACM-Reference-Format}
\bibliography{Reference}

\end{document}